\documentstyle[pre,aps,multicol,epsf]{revtex}
\renewcommand{\narrowtext}{\begin{multicols}{2} \global\columnwidth20.5pc}
\renewcommand{\widetext}{\end{multicols} \global\columnwidth42.5pc}

\begin{document}
\draft
\title{Parametric level statistics in random matrix theory: Exact solution}
\author{E. Kanzieper}
\address{The Abdus Salam International Centre for Theoretical Physics, P.O.B. 586, 34100 Trieste, Italy}
\date{November 18, 1998}
\maketitle

\begin{abstract}
An exact solution to the problem of parametric level statistics in non--Gaussian ensembles
of $N \times N$ Hermitian random matrices with either soft or strong level confinement
is formulated within the framework of the orthogonal polynomial technique. Being applied 
to random matrices with strong level confinement, the solution obtained leads
to emergence of a new connection relation that makes a link between the parametric level statistics
and the scalar two--point kernel in the thermodynamic limit.
\end{abstract}


\narrowtext

In recent years, the theory of {\it non--Gaussian} random matrix ensembles 
has experienced a sound progress motivated by new ideas \cite{GMGW-1998} in quantum
chromodynamics (QCD) and mesoscopic physics. Both fields, concentrating on physically
different objects which besides have incommensurable spatial scales, make use of the
same language of random matrix theory \cite{M-1991} to reveal the universal statistical 
manifestations of symmetries inherent in a physical system. In QCD, invariant non--Gaussian 
random matrix models incorporating global symmetries of the QCD Dirac operator appear
to describe the universal features \cite{V-1998} of low--energy part of the spectrum of 
the Dirac equation. In mesoscopic physics, similar matrix models of appropriate symmetries 
arise in identifying the universal electron level statistics in 
normal--superconducting hybrid structures \cite{AZ-1997}, as well as in the 
context of mesoscopic electron transport \cite{NS-1993}. A new class of non--Gaussian 
random matrices \cite{MCIN-1993}
with finite level compressibility should also be mentioned; these serve 
as a basis for constructing 
the toy models \cite{KM-1997} of universal spectral statistics expected at the edge of the 
metal--insulator transition.

One of the most promising ways of studying the non--Gaussian random matrix ensembles
consists of revealing the universal connection relations (i) between different entities 
(statistics) within the same matrix model, and (ii) between the same statistics in 
different matrix models (which only differ by underlying symmetries). In both cases the 
connection identities (if they exist)
originate from underlying mathematical structure of random matrices which deserves to
be studied in its own right, also. Being brought together, these two approaches have led to
a substantial breakthrough in understanding the universality phenomenon in invariant 
random matrix ensembles characterized by strong level confinement \cite{Strong}.

(i) The Shohat method of Refs. \cite{KF-19978,KF-1999} is a
typical example of the first approach. This formalism, employing the free fermion representation
of the joint distribution function of eigenvalues in unitary invariant random matrix ensembles,
reduces the problem of finding the $n$--point spectral correlators in the thermodynamic limit
to a simpler problem of solving an effective one--particle Schr\"odinger equation. In the 
absence of a fine tuning of confinement potential to so--called multicritical points 
\cite{ADMN-1998}, this equation involves two ingredients: 
the macroscopic mean density of states $\nu_N(E)$ (which is a {\it one--point} spectral characteristics) 
and the term determined by the singularities of confinement potential. In this way, one relates
the $n$--point spectral characteristics to the one--point characteristics.

(ii) Recent studies by \c Sener and Verbaarschot \cite{SV-1998}, and by Widom \cite{W-1998} are 
representatives of the second approach. These authors, by using the two completely different 
constructions (of skew \cite{BN-1991} and ordinary \cite{TW-1998} orthogonal polynomials, 
respectively) succeeded in deriving universal relations between eigenvalue correlations in 
a class
of non--Gaussian random matrix ensembles with the same, finite polynomial, confinement potential
but with different symmetries. These remarkable relations express the spectral correlation
functions of matrix models with orthogonal $(\beta=1)$ and symplectic $(\beta=4)$ symmetries 
in terms of the scalar
two--point kernel of the corresponding ensemble with unitary symmetry $(\beta=2)$.

Contrary to conventional spectral statistics, which has received a detailed
study in the literature, quite a bit is known \cite{BZ-1994,HW-1995} about parametric 
level statistics in non--Gaussian random matrix models. In the present Letter, we address
this problem for ensembles with unitary symmetry, simultaneously with taking a further
step in establishing the connection relations in non--Gaussian random matrix theory. Previously, 
the issue of parametric 
correlations in non--Gaussian random matrix ensembles (on the scale of several mean level 
spacings) has been considered only by Hackenbroich 
and Weidenm\"uller \cite{HW-1995} who used a supersymmetry formalism to provide a 
heuristic proof that in all three symmetry classes $(\beta=1,2$ and $4)$ the parametric 
spectral correlations of all orders 
in invariant matrix model given by Eq. (\ref{jdf}) below are insensitive to 
the particular form of the confinement potential $v[{\bf H}_0]$, and thus
follow the predictions found for Gaussian ensembles. This statement is valid for strong level 
confinement and refers to the bulk of the spectrum. Until now, however, no {\it constructive}
formalism was proposed to quantitatively describe parametric level statistics in a random
matrix ensemble with a general non--Gaussian measure. Our aim here is to formulate an {\it exact
solution} to the problem of parametric level statistics for $\beta=2$. Our study differs
from Refs. \cite{BZ-1994,HW-1995} in three main respects: (i) It applies to both strong {\it and} 
soft level confinement; (ii) it is valid in arbitrary spectrum range; (iii) it gives a prescription
of how to {\it actually} compute parametric statistics for both arbitrary matrix size $N$
and in the thermodynamic limit. As an illustration, we apply the formalism developed
to obtain closed analytical expressions for parametric eigenlevel correlations
in the model Eq. (\ref{jdf}) for strong level confinement in the origin, the bulk, and
the soft--edge scaling limits \cite{KF-1999}.

Now we turn to the formalization of the problem and to the statement of the main results. 
Consider 
an unperturbed matrix ensemble defined by the joint distribution function of the entries of 
$N\times N$ Hermitian matrix ${\bf H}_0$, 
\begin{equation}
{\cal P}_1\left[ {\bf H}_0\right] ={\cal Z}_N^{-1}\exp \left\{ -
\mathop{\rm Tr}
v\left[ {\bf H}_0\right] \right\} .  \label{jdf}
\end{equation}
Here $v\left[ {\bf H}_0\right] $ is an arbitrary (not necessarily strong) confinement potential
ensuring existence of the partition function ${\cal Z}_N$. 
Eigenlevel correlations of arbitrary order in this ensemble are known \cite{M-1991} to be expressed
through the scalar two--point kernel $K_N\left( E, E^{\prime}\right) = 
\sum_{n=0}^{N-1}\psi _n\left( E\right) \psi _n\left( E^{\prime}\right)$. Here the 
fictitious wave functions $\psi _n\left( E\right) = {\cal N}_n^{-1/2}\exp \left\{-v\left(E\right)/2\right\} p_n\left(E\right)$
involve a set of polynomials $p_n\left(E\right)$ orthogonal on the entire real
axis {\bf R} with respect to the
measure $d\mu(E)=\exp \left\{-v\left(E\right)\right\}dE$, 
\begin{equation}
\int d\mu\left(E\right) p_n\left(E\right)p_m\left(E\right)={\cal N}_n
\delta_{nm}. \label{normal}
\end{equation}
In the presence of the perturbation, which we parameterize by the strength $\phi$,
the Hamiltonian of the system modeled by the matrix ${\bf H}_0$ flows to ${\bf H}_\phi ={\bf H}_0+\phi {\bf H}_1$.
We choose the perturbation matrix ${\bf H}_1$ to belong to the Gaussian Unitary 
Ensemble (GUE), $\delta {\cal P}\left[ {\bf H}_1\right] \propto \exp \left\{ -\mathop{\rm Tr}{\bf H}_1^{2} \right\} $, 
and assume $\phi $ to be positive without any loss of generality. Under these
assumptions, the connected part of the parametric level densities
correlator defined as
\begin{eqnarray}
R_N^{(c)}\left( E,E^{\prime };\phi \right) &=&\left\langle 
\mathop{\rm Tr}\delta \left( E-{\bf H}_0\right) \mathop{\rm Tr}\delta \left( E^{\prime
}-{\bf H}_\phi \right) \right\rangle \nonumber
\\
&-&\left\langle \mathop{\rm Tr}%
\delta \left( E-{\bf H}_0\right) \right\rangle \left\langle \mathop{%
\rm Tr}\delta \left( E^{\prime }-{\bf H}_\phi \right) \right\rangle,
\label{eq03}
\end{eqnarray}
where the angular brackets $\left\langle \ldots \right\rangle$ stand for
averaging over the ensembles of random matrices ${\bf H}_0$ and ${\bf H}_1$, is
given by
\begin{eqnarray}
R_N^{(c)}&&\left( E,E^{\prime };\phi \right) \equiv -\frac 1\pi \int_{-\infty }^{+\infty
}d\sigma _1 \int_{-\infty }^{+\infty
}d\sigma _2
\mathop{\rm e}{}^{-(\sigma _1^2+\sigma _2^2)} \nonumber \\ 
&&\times 
\mathop{\rm e}{}^{\frac 12[v(E^{\prime }+i\phi \sigma _1)-v(E^{\prime }+\phi
\sigma _2)]}
K_N\left( E,E^{\prime }+i\phi \sigma _1\right) \nonumber \\ 
&&\times \left[
K_N\left( E^{\prime }+\phi \sigma _2,E\right) -\delta \left( E^{\prime
}+\phi \sigma _2 - E\right) \right]. 
\label{eq04}
\end{eqnarray}
Equation (\ref{eq04}) is {\it exact} for arbitrary finite $N$, and the 
sum rule for $R_N^{(c)}$ is fulfilled by construction. Note, that for 
finite $N$, the equation contains both the two--point kernel 
$K_N\left( E,E^{\prime }\right)$ of an unperturbed matrix ensemble Eq. (\ref{jdf}) 
and the confinement potential $v(E)$. It is worth mentioning that only for the GUE measure, $v[{\bf H}_0] 
\propto {\bf H}_0^2$, this finite--$N$ 
solution reduces to the one known from the Dyson Brownian motion model \cite{BR-1994} 
in the limit $\phi \ll 1$. 

Thermodynamic limit $N \rightarrow \infty$ of Eq. 
(\ref{eq04}) is substantially different in the case of the soft and the strong level confinement.
(i) If the confinement potential is {\it soft}, the two--point kernel $K_N$ has a well defined large--$N$
limit given by a corresponding Poisson kernel $K_P(E, E^{\prime};r)=\sum_{n=0}^{\infty} r^n \psi_n (E) \psi_n (E^{\prime})$
 taken at $r=1$ \cite{Remark3}. As a consequence, Eq. (\ref{eq04}) holds in the large--$N$ limit as it stands,
with $K_N$ replaced by $K_P(r=1)$. After implementation of $N$--independent unfolding, $E \mapsto s$, $\phi \mapsto X$, it
provides an exact answer. Hence, generically there are no reasons 
to expect a universality of parametric level statistics in ensembles with soft level confinement. 
(ii) For {\it strong} level confinement, Eq. (\ref{eq04}) can take a universal 
form in the thermodynamic limit after the implementation
of the unfolding procedure $ds/dE=\overline{\nu}_N$, $X=\phi \overline{\nu}_N$ (with $\overline{\nu}_N$ being the local 
level density \cite{Unfold}),
\begin{eqnarray}
R^{(c)}&&\left( s,s^{\prime };X \right) = -\frac 1\pi \int_{-\infty }^{+\infty
}d\sigma _1 \int_{-\infty }^{+\infty
}d\sigma _2
\mathop{\rm e}{}^{-(\sigma _1^2+\sigma _2^2)} \nonumber \\
&&\times F_X\left(s^{\prime}, \sigma_1, \sigma_2\right)
K\left( s,s^{\prime }+iX \sigma _1\right) \nonumber \\ 
&&\times \left[
K\left( s^{\prime }+X \sigma _2,s\right) -\delta \left( s^{\prime}+X \sigma _2-s\right) \right], 
\label{eq05}
\end{eqnarray}
where
\begin{eqnarray}
F_X\left(s,\sigma_1, \sigma_2\right)&& = \lim _{N\rightarrow \infty } 
\frac {\exp \{ \frac 12[v(E_{N,s}+i\phi_{N,X} \sigma_1)]\}} 
{\exp \{ \frac 12 [v(E_{N,s}+\phi_{N,X} \sigma_2)]\}}.
\label{lim}
\end{eqnarray}
Equations (\ref{eq05}) and (\ref{lim}) establish a link between the parametric level statistics in
unitary invariant non--Gaussian ensemble of random matrices and the energy
level statistics in the same, unperturbed random matrix ensemble, through
the unfolded two--point kernel 
\begin{eqnarray}
K\left( s,s^{\prime }\right) &&\equiv 
\lim_{N \rightarrow \infty} K_N(E_{N,s},
E_{N,s^{\prime}}) \nonumber \\
&&\times {\nu_N^{-1/2}(E_{N,s}) \nu_N^{-1/2}(E_{N,s^{\prime}}) }.
\label{unker}
\end{eqnarray}
Connection relations Eqs. (\ref{eq04}) -- (\ref{unker}) provide a constructive 
rule for computing parametric level statistics in non--Gaussian matrix models and
constitute the main results of the paper.

In particular, when {\it both} $F_X$ and $K\left( s,s^{\prime }\right)$ are universal functions 
for a class of matrix models, the universality
of eigenlevel correlations will automatically imply the universality
of parametric level correlations. To demonstrate this point, consider the matrix model Eq. (\ref{jdf})
with strong confinement potential exhibiting a logarithmic singularity, 
$v(E) \rightarrow v(E)-2\alpha \log |E|$ $(\alpha > -1/2)$. For $\beta=2$ and in the absence of the perturbation,
this model has been proven to possess three different types of locally universal
eigenlevel correlations in the bulk (BSL) \cite{BZ-1993}, the origin (OSL) 
\cite{ADMN-1997,KF-19978} 
and the soft--edge (SSL) scaling limits \cite{KF-19978}. In all
these scaling limits, the two--point kernel obeys the universal laws 
which we convert to
\begin{eqnarray}
K(s,s^{\prime}) = \frac{\Psi_1(s)\Psi_2(s^{\prime}) - \Psi_2(s)\Psi_1(s^{\prime})}{s-s^{\prime}}.
\label{kerker}
\end{eqnarray}
Here $s$ and $s^{\prime}$ are appropriately unfolded spectral variables, 
\begin{equation}
\label{a1}\Psi _1\left( s\right) =\left\{ 
\begin{array}{cc}
\pi ^{-1}\sin \left( \pi s\right) , & \text{BSL,} \\ -(
s/2\pi)^{1/2}J_{\alpha -1/2}\left( \pi s\right) , & \text{OSL,}
\\ \mathop{\rm Ai}\left( s\right) , & \text{SSL,}
\end{array}
\right.
\end{equation}
$\Psi _2\left( s\right) =\Psi _1^{\prime }(s)$ for BSL and SSL, while for
OSL $\Psi _2\left( s\right) =\Psi _1^{\prime }(s) - \alpha s^{-1} \Psi_1 (s)$. 
These three universal kernels are 
referred to as the sine, the Bessel, and the Airy kernels, respectively. 
Universal parametric level statistics in this ensemble is a consequence
of the connection relations Eqs. (\ref{eq05}) and (\ref{lim}), which enable 
us to write down closed analytical expressions for the parametric level density 
correlator $R^{(c)}$ just by noting 
that $F_X$ is a universal function: $F_X(s,\sigma_1, \sigma_2) = 
{\left( s+ X \sigma_2\right)}^{\alpha} {\left( s+ i X \sigma_1\right)}^{-\alpha}$
in the origin scaling limit, and $F_X = 1$ otherwise.

Let us now outline the derivation of the announced relations. The two--point 
parametric correlator $R_N\left( E, E^{\prime };\phi \right)$
of level densities can be represented as an average over the joint distribution function
\begin{equation}
{\cal P}_2\left[ {\bf H}_0, {\bf H}_\phi \right] \propto \exp \left\{ -
\mathop{\rm Tr}
\left( v\left[ {\bf H}_0\right] +\phi ^{-2}({\bf H}_0-{\bf H}_\phi )^2\right) \right\} ,
\label{pjdff}
\end{equation}
of two Hamiltonians, unperturbed ${\bf H}_0$ and perturbed ${\bf H}_\phi$,
\begin{equation}
R_N\left( E, E^{\prime };\phi \right) =\left\langle 
\mathop{\rm Tr}
\delta \left( E -{\bf H}_0\right) 
\mathop{\rm Tr}
\delta \left( E^{\prime }-{\bf H}_\phi \right) \right\rangle _{{\cal P}_2}.
\label{pldc}
\end{equation}
In Eq. (\ref{pldc}) the averaging over the distribution ${\cal P}_2$ can explicitly 
be performed by using the Eynard--Mehta theorem \cite{EM-1998}. The theorem states 
that $R_N\left(E, E^{\prime };\phi \right)$ equals the determinant of the $%
2\times 2$ matrix kernel, 
\begin{equation}
R_N\left( E, E^{\prime };\phi \right) =
\mathop{\rm Det}
\left[ k_{\alpha \beta }^{(N)}\left( E_\alpha ,E_\beta \right) \right] _{\alpha
,\beta =1,2},  \label{emth}
\end{equation}
where $E_\alpha =E +\left( E^{\prime }-E \right)
\left[ 1+\left( -1\right) ^\alpha \right] /2$. An explicit representation of
the matrix kernel reads, 
\begin{eqnarray}
k_{\alpha \beta }^{(N)}\left( \xi, \xi^{\prime }\right)  &=&\phi
^{-1}\sum_{n=0}^{N-1}h_n^{-1}P_{\alpha ,n}\left( \frac \xi \phi
\right) Q_{\beta ,n}\left( \frac{\xi^{\prime }}\phi \right)   \nonumber
\\
&&-\phi ^{-1}\delta _{\alpha ,\beta -1}W\left( \frac \xi \phi ,\frac{%
\xi^{\prime }}\phi \right) .  \label{ker2}
\end{eqnarray}
Here $P_{\alpha ,n}$ and $Q_{\alpha ,n}$ are four complete sets of
orthogonal functions; two of them, $P_{1,n}=P_n$ and $Q_{2,n}=Q_n$, are
basic polynomials orthogonal with respect to the {\it nonlocal} weight $W\left(
\xi ,\eta \right) =\exp \left\{ -v(\phi \xi )-(\xi -\eta )^2\right\} $, 
\begin{equation}
\int d\xi \int d\eta P_n\left( \xi \right) W\left( \xi ,\eta \right)
Q_m\left( \eta \right) =h_n\delta _{nm}
\label{ops}
\end{equation}
with $h_n$ being a normalization constant. (Hereafter, integration over single variable 
runs over entire real axis). Two
remaining orthogonal functions are connected to the basic polynomials as
follows:
\begin{mathletters}
\label{of}
\begin{eqnarray}
P_{2,n}\left( \xi \right)  &=&\int d\eta P_n\left( \eta \right) W\left( \eta
,\xi \right) ,  \label{of.1} \\
Q_{1,n}\left( \xi \right)  &=&\int d\eta W\left( \xi ,\eta \right) Q_n\left(
\eta \right).  \label{of.2}
\end{eqnarray}
\end{mathletters}
\noindent
Remarkably, all of the four sets $P_{\alpha ,n}$ and $Q_{\alpha ,n}$ can solely be expressed
through the orthogonal polynomials $p_n(E)$, Eq. (\ref{normal}), which determine the eigenlevel statistics 
in an unperturbed random matrix ensemble Eq. (\ref{jdf}). To prove this, we use the 
representation \cite{E-1998} of polynomials $P_n$ and $Q_n$ orthogonal with respect to a 
nonlocal measure $W$ as a determinant averaged over the distribution 
${\cal P}_2\left[ \phi {\bf M}_1,\phi {\bf M}_2\right] $, where ${\bf M}_1$ and ${\bf M}_2$ 
are $n\times n$ Hermitian matrices: 
\begin{mathletters}
\label{ef}
\begin{eqnarray}
P_n\left( \xi \right)  &=&\left\langle 
\mathop{\rm Det}
_{n\times n}\left( \xi -{\bf M}_1\right) \right\rangle _{{\cal P}_2\left[ \phi
{\bf M}_1,\phi {\bf M}_2\right] },  \label{mp} \\
Q_n\left( \xi \right)  &=&\left\langle 
\mathop{\rm Det}
_{n\times n}\left( \xi -{\bf M}_2\right) \right\rangle _{{\cal P}_2\left[ \phi
{\bf M}_1,\phi {\bf M}_2\right] }.  \label{mq}
\end{eqnarray}
\end{mathletters}
\noindent
Integration over matrix ${\bf M}_2$ in Eq. (\ref{mp}) is straightforward; it yields
the expression
\begin{eqnarray}
P_n\left( \xi \right)  =\left\langle 
\mathop{\rm Det}
_{n\times n}\left( \xi -{\bf M}_1\right) \right\rangle _{{\cal P}_1\left[ \phi
{\bf M}_1\right] }=\phi^{1/2} p_n\left(\phi \xi\right),  \label{pk} 
\end{eqnarray}
which we have recognized as a matrix--integral representation \cite{S-1967} of the
ordinary orthogonal polynomials appearing in Eq. (\ref{normal}). To integrate 
out the matrix ${\bf M}_2$ in Eq. (\ref{mq}) we rewrite the determinant there in the form of
the integral $\mathop{\rm Det}
_{n\times n}\left( \xi -{\bf M}_2\right) = \int D\left[\chi\right] \exp\left\{-\chi^{\dagger}\left(
\xi - {\bf M}_2\right) \chi \right\}$ over $n$--component Grassmann vector
$\chi = (\chi_1,\ldots,\chi_n)$. Further integration over ${\bf M}_2$ leads to
\begin{eqnarray}
Q_n\left(\xi\right) &=&  \int D\left[\chi\right] 
\exp \{-\frac{1}4 (\chi^{\dagger} \chi )^{2}\} \nonumber \\
&&\times \left\langle \exp\left\{-\chi^{\dagger}\left(
\xi-{\bf M}_1\right)\chi\right\} \right\rangle _{{\cal P}_1\left[ \phi
{\bf M}_1\right] }.  \label{qkg}
\end{eqnarray}
Decoupling the `interaction' term $\exp \{- (\chi^{\dagger} \chi )^{2}/4\}$ in 
Eq. (\ref{qkg}) followed by integration over the Grassmann vector $\chi$ results in
\begin{eqnarray}
Q_n \left(\xi\right) = \pi^{-1/2} \int d\sigma_1 \mathop{\rm e}{}^{-\sigma _1^2}
P_n\left(\xi + i\sigma_1\right). \label{qkg2}
\end{eqnarray}
Combining Eqs. (\ref{of}), (\ref{pk}) and (\ref{qkg2}), we obtain
explicit expressions for $P_{\alpha ,n}$ and $Q_{\alpha ,n}$ entering Eq. (\ref{ker2})
for the matrix kernel $k_{\alpha \beta }^{(N)}$, 
\begin{mathletters}
\label{sets}
\begin{eqnarray}
P_{1,n}\left( \xi \right)  &=&\phi ^{1/2}p_n\left( \phi \xi \right) ,
\label{set1} \\
P_{2,n}\left( \xi \right)  &=&\phi ^{1/2}\int d\eta W\left( \xi ,\eta
\right) p_n\left( \phi \eta \right) ,  \label{set2} \\
Q_{1,n}\left( \xi \right)  &=&\left( \pi \phi \right) ^{1/2}
\mathop{\rm e}
{}^{-v(\phi \xi )}p_n\left( \phi \xi \right) ,  \label{set3} \\
Q_{2,n}\left( \xi \right)  &=&\left( \pi ^{-1}\phi \right) ^{1/2}\int
d\sigma_1
\mathop{\rm e}
{}^{-\sigma_1 ^2}p_n\left( \phi (\xi +i\sigma_1 )\right) .  \label{set4}
\end{eqnarray}
\end{mathletters}
\noindent
One can verify that the orthogonality relation Eq. (\ref{ops}) is
fulfilled, and the normalization constants in Eqs. (\ref{normal}) and (\ref{ops})
are related as $h_n=\pi^{-1/2}{\cal N}_n$. Substituting Eqs. (\ref{sets}) into 
Eqs. (\ref{ker2}) and (\ref{emth}), and subtracting disconnected part given by the product 
$k_{11}^{(N)} k_{22}^{(N)}$ is the final stage of the calculations which leads to
our solution Eq. (\ref{eq04}). Equations (\ref{eq05}) and (\ref{lim}) follow
upon appropriate unfolding. Note that these results can be extended to the 
$n$--point parametric correlators along the lines of Ref. \cite{KF-1998}.

In conclusion, we presented an exact solution, Eq. (\ref{eq04}), to the problem of parametric 
level statistics in unitary invariant, non--Gaussian random matrix ensembles characterized 
by either soft or strong level confinement. The solution is exact in precisely the same 
sense as are the formulas by Gaudin and Mehta \cite{M-1991} for conventional spectral 
statistics. Generically, the parametric `density--density' correlator was shown to 
depend on both the two--point scalar kernel and the level confinement, through a 
double integral transformation which, in turn, provides a constructive tool for description 
of parametric level correlations in non--Gaussian random matrix theory. In random matrix 
ensembles with strong level confinement, the solution presented takes a particular simple 
form in the thermodynamic limit leading to emergence of a new connection relation, 
Eq. (\ref{eq05}), between the parametric spectral statistics and the scalar two--point
kernel of an unperturbed ensemble. In the case of soft level confinement, the formalism
developed is potentially applicable to a study of parametric level statistics in unitary 
invariant random matrix model with log--squared confinement potential \cite{MCIN-1993}, 
$v(E) \propto \log^2 |E|$ at $|E| \gg 1$. The latter problem is of conceptual interest in 
view of the recent work \cite{KM-1997} where existence of a new class of random matrix 
ensembles with finite eigenlevel compressibility and multifractal eigenvectors was 
suggested. There, identical two--level correlation functions in proper regimes in three 
different ensembles -- (a) generalized matrix model of
Ref. \cite{MNS-1994}, (b) power--law banded matrices of Ref. \cite{MFDQS-1996}, and (c) 
rotationally invariant random matrices with log--squared confinement potential already 
mentioned above -- were one of the key points to identifying the new universality class 
relevant to the one expected at the edge of the metal--insulator transition. In this context, 
it would be desirable to learn whether these three matrix models also enjoy the same 
{\it parametric} level statistics. At least for two of the ensembles [(b) and (c)]
such a verification seems now to be possible.

I thank V. E. Kravtsov, K. A. Muttalib and I. Yurkevich for discussions at different
stages of this study. 

\widetext

\begin{references}
\bibitem{Strong} The confinement potential $v(E)$ in Eq. (\ref{jdf}) is said to be `strong', if it
increases at least as $\left| E \right|$ at infinity; otherwise it is referred to
as `soft' level confinement.
\end{references}
\end{document}